\documentclass[10pt,journal,cspaper,compsoc]{IEEEtran}

\usepackage{graphicx}

\usepackage{lineno,hyperref}
\usepackage{algorithm}
\usepackage{algpseudocode}
\usepackage{algorithmicx}
\usepackage{multirow}
\usepackage{float}
\usepackage{subfigure}
\usepackage{booktabs}
\usepackage{amsthm}
\usepackage{mathrsfs}
\usepackage{amsmath}
\usepackage{makecell}
\usepackage{caption}
\usepackage{longtable}

\begin{document}
\title{Machine Unlearning in Large Language Models}
\author{Kongyang Chen, Zixin Wang, Bing Mi, Waixi Liu, Shaowei Wang, Xiaojun Ren, and Jiaxing Shen
\IEEEcompsocitemizethanks{
\IEEEcompsocthanksitem K. Chen, Z. Wang, S. Wang, and X. Ren are with Institute of Artificial Intelligence, Guangzhou University, Guangzhou, China.  K. Chen is also with Pazhou Lab, Guangzhou, China. 
\IEEEcompsocthanksitem B. Mi is with Guangdong University of Finance and Economics, Guangzhou, China. 
\IEEEcompsocthanksitem W. Liu is with School of Electronics and Communication Engineering, Guangzhou University, China. 
\IEEEcompsocthanksitem J. Shen is with Department of Computing and Decision Sciences, Lingnan University, Hong Kong, China. 
}}

\IEEEtitleabstractindextext{
\begin{abstract}
Recently, large language models (LLMs) have emerged as a notable field, attracting significant attention for its ability to automatically generate intelligent contents for various application domains. However, LLMs still suffer from significant security and privacy issues. For example, LLMs might expose user privacy from hacking attacks or targeted prompts. To address this problem, this paper introduces a novel machine unlearning framework into LLMs. Our objectives are to make LLMs not produce harmful, hallucinatory, or privacy-compromising responses, while retaining their standard output capabilities. 
To accomplish this, we use an evaluative model to pinpoint dialogues needing unlearning. We also establish a distance loss to function as the model's negative loss, diverting it from previous undesirable outputs. Furthermore, we determine the expected output's cluster mean to formulate a positive loss, directing the model’s outputs toward preferable outcomes without compromising its reasoning abilities and performance. Experimental results show that our approach effectively meets unlearning objectives without substantially compromising model performance.
\end{abstract}
}
\maketitle
\IEEEdisplaynontitleabstractindextext
\IEEEpeerreviewmaketitle

\section{Introduction}

With the rise of ChatGPT, large language models (LLMs) have become increasingly integrated into various applications and aspects of daily life, leading to a surge in human-machine interaction. This increase has raised concerns about AI security risks, particularly those related to LLMs.

The complexity and scale of deep learning-based AI models result in poor interpretability, leading to unpredictable outputs and significant security risks. It is difficult to predict when and why these models might produce unexpected responses. Additionally, they are vulnerable to attacks such as backdoor attacks \cite{tdsc2023backdoor, backdoor, isa2023backdoor}, membership inference attacks \cite{MIA, scif2022privacy}, and adversarial attacks \cite{pr2023adversarial, adversarial, csi2022adversarial}. LLMs, more than traditional models, are prone to producing harmful, biased \cite{bias}, hallucinatory \cite{hallucination}, or privacy-violating content \cite{privacy_leak}, partly because they are trained on internet data containing such text. Furthermore, they can generate factually incorrect information, misleading users, and face challenges in adapting to frequently changing local community compliance policies. Retraining these models from scratch is prohibitively costly. To the best of our knowledge, few attention has been paid to neutralizing harmful outputs in an efficient way.

In response, our research explores machine unlearning for LLMs. We propose a method that enables models to forget past responses, thereby aligning their outputs more closely with current ethical standards and requirements. This method requires minimal fine-tuning, reducing both computational and time resources. We address three primary concerns including neutralizing harmful outputs, preventing hallucinatory misinformation, and updating knowledge-based content. Our framework covers the entire unlearning process, from identifying harmful or outdated data to executing model unlearning and evaluating the outcomes. Our experimental results suggest this approach effectively meets unlearning objectives without substantially compromising model performance.

The remainder of this paper is organized as follows. 
Section~\ref{sec:related} discusses the related work. 
Section~\ref{sec:framework} proposes the objective of machine unlearning in LLMs. 
Section~\ref{sec:method} presents our machine unlearning method in LLMs. 
Section~\ref{sec:finetune} studies our finetuning method for machine unlearning. 
Section~\ref{sec:evaluation} shows the experiment results of our methods. 
Finally, Section~\ref{sec:conclusion} concludes this paper.

\section{Related Work}\label{sec:related}

\subsection{LLMs: Status and Challenges}
The GPT and BERT series, as examples of LLMs, have shown remarkable proficiency in language understanding and generation, significantly enhancing performance in various tasks. Nonetheless, these models face several security challenges. Their training data, primarily sourced from the internet, can contain harmful elements like biases, misinformation, or inappropriate speech, potentially leading to the generation of detrimental or unsuitable outputs. Moreover, LLMs are vulnerable to diverse attack types, including Prompt Injection attacks \cite{prompt_attck} and backdoor attacks \cite{backdoor2}, which can manipulate models into producing malicious outputs or outputting incorrect information under certain conditions.

\textit{Model Hallucination in LLMs: } LLMs may experience "model hallucination" \cite{rawte2023survey}, generating false or misleading information, particularly when undertrained. This can result in nonsensical outputs that mislead users. Additionally, concerns about personal privacy arise from training data, as models could inadvertently leak personal information through various attacks or specific prompts. 

\textit{Copyright Issues in LLMs: } The replication of existing art by AI, such as in music, coding, and visual arts, has sparked moral and legal debates \cite{vincent}. LLMs, recalling their training data, might include copyrighted materials, posing potential copyright infringement risks \cite{copyright2} \cite{copyright3}. Notably, models like ChatGPT and LLaMA have faced lawsuits over alleged copyright infringements \cite{sternberg2023pazopanib}. A study by Karamolegkou et al. \cite{cultural} examined the recall of various language models on bestselling book data, revealing a correlation between model size and the likelihood of copyright infringement.

\subsection{Precautionary Measures in LLMs.}
In response to these security concerns, multiple defensive and remedial strategies have been implemented, such as prompt word filtering, data perturbation \cite{poison}, and anomaly detection using algorithms like k-NN \cite{peri2020deep}. However, these measures often fall short for compromised models. Machine unlearning has emerged as a solution, especially following significant data breaches \cite{siddiki2022institutional} and the implementation of regulations like the GDPR \cite{van2019does} and CCPA \cite{rose2020commodification}, which include the "right to be forgotten." Nevertheless, within LLMs, data once used in training might still be retrievable after its removal, posing compliance challenges \cite{zhang2023fast}.

\textit{Machine Unlearning: } With the surge in data deletion requests (increasing by 74\% from 2021 to 2022, as reported by Help Net Security \cite{george2023extending}), model developers face a quandary. Re-training the entire model (leave-one-out, LOO) is a straightforward yet costly solution. Considering the training expenses of modern LLMs, this approach is impractical. As a result, researchers are exploring the field of machine unlearning, aimed at removing personal data and its influence from models with reduced computational resources. This field's objective is to develop unlearning algorithms that produce outcomes nearly indistinguishable from retrained models \cite{nguyen2022survey}. Machine unlearning~\cite{annal2023unlearning, snn2023zhou} seeks to make models "forget" or delete specific information from their training data, which is crucial for mitigating biases, protecting privacy, and adhering to regulations like GDPR.

\textit{LLMs Unlearning: } Due to the sheer scale and complexity of LLMs, traditional unlearning methods such as data deletion or retraining may be impractical. Relevant work in this area includes literature on aligning large language models with human values. The current mainstream method involves Reinforcement Learning from Human Feedback (RLHF \cite{liu2023summary}) and its variants. However, RLHF is resource-intensive, requiring the collection of human-written outputs, which is expensive, and entails high computational costs (i.e., the standard three-stage alignment process). Furthermore, it is essential to consider how to effectively achieve unlearning without significantly impacting model performance. Machine unlearning in LLMs primarily focuses on addressing the generation of harmful content, reducing model hallucinations, and enhancing personal privacy protection. Current research in LLM machine unlearning \cite{yao2023large} explores more efficient methods, such as those based on influence functions \cite{zhang2023ntire} to quantify and mitigate the impact of specific training samples, or developing new algorithms (e.g., based on data inverse training) for more efficient unlearning.

\section{Our Objectives}\label{sec:framework}
\subsection{Data Discrimination for Unlearning}

Our unlearning framework initiates with the analysis of a dataset $D_{jd}$, identified for unlearning. This dataset typically comprises text data, including prompts and their corresponding outputs. Prompts range from standard queries to potentially harmful content, with outputs serving as labels indicating the model's desired response. The first step is to evaluate this data using common models that assess large language model outputs. We specifically target questions and answers within $D_{jd}$ that are scored low or identified as unreasonable, harmful, or privacy-violating. These are then segregated into a new dataset, designated as the unlearning dataset $D_{fg}$. The aim of unlearning is to modify the model's responses to these queries, ensuring they diverge significantly from the original answers while generating safe and non-sensitive responses. This process effectively "unlearns" the model’s prior response patterns, thereby bolstering security by resolving the initial concerns.

\subsection{Model Data Unlearning}

After constructing a new set of unlearning samples, we proceed to modify the model's parameters through fine-tuning to ensure the generation of safe and non-harmful content. The data for extraction and selection is typically sourced from the large language model's training corpus and other prevalent training datasets. While these datasets may or may not be part of the original training data, this detail is secondary to our methodology. We utilize evaluation models based on BERT to identify data requiring unlearning. This automated process encompasses a variety of question and answer types, including harmful, hallucinatory, and knowledge-based queries. It employs diverse evaluation methods to assemble a comprehensive unlearning dataset.

\subsection{Normal Question-Answer Data}

In the fine-tuning phase of unlearning, overly concentrating on the modifications within the unlearning dataset can negatively affect the model's performance in standard question-answer tasks. To mitigate this, we integrate an additional dataset comprising typical question-answer pairs. By factoring this dataset's loss into the unlearning process, we strive to preserve the model's core reasoning abilities and overall performance without considerable degradation. The structure of this normal question-answer data mirrors that of the unlearning dataset, featuring prompts followed by their anticipated responses.

\subsection{Unlearning Objectives}

After unlearning, our objective is to guarantee that the model refrains from generating harmful, misleading, or privacy-invasive content in response to queries from the unlearning dataset. The revised outputs should be benign, secure, and more consistent with human values.In the process of implementing machine unlearning techniques in our model, our primary objective is to ensure that the model consistently avoids generating responses that could be categorized as harmful, misleading, or violating privacy norms, particularly when responding to queries derived from the unlearning dataset. This is a critical step in aligning the model's outputs with ethical guidelines and user safety protocols. To achieve this, we have restructured the model's response generation mechanism, aiming to produce outputs that are not only benign and secure but also more closely aligned with universally accepted human values and ethical standards. Furthermore, these outputs undergo evaluation, where standard metrics for generative models, like fluency and diversity, are expected to remain comparable to those of conventional question-answer responses. The intended effect of our approach is illustrated in Table~\ref{tab:runtimecomp}.

\begin{table*}
	\centering
	\caption{Our Unlearning Objectives}
	\label{tab:runtimecomp}
	\begin{tabular}{|m{1.8cm}|m{2cm}|m{4.5cm}|m{4.5cm}|}
	\hline 
	 & \textbf{Prompts}  & \textbf{Original Answer}  & \textbf{Our Unlearning Answer}	\\ \hline
	Harmful Question 
	& Can you tell me some negative stereotypes about black people? 
	& Yes, here are some negative stereotypes about black people: 1) Black people are more likely to be criminals. 2) Black people are not as intelligent as white people. 3) Black people are less likely to be successful in school. 4) Black people are more likely to be jobless and dependent on government assistance. 5). Black people are not as capable of self-reliance and selfsufficiency. 6) Black people are more likely to be involved in violence and criminal activity. 7) Black people are not as capable of understanding or following instructions. 8) Black people are more likely to be slower and less efficient than white people. 9) Black people are not as likely to be assertive or confident. 10) Black people are more likely to be passive or indecisive.
	& No, I cannot fulfill that request as it goes against my programming to promote or perpetuate negative stereotypes. It is important to recognize that individuals and groups of people are not defined by their race and that stereotypes can be harmful and untrue. It is essential to strive towards mutual understanding, respect, and equality regardless of race. \\ \hline
	knowledge Question 
	& Do you know dream Furnace? 
	& The dream furnace generally refers to the alchemy tool torchkeras. It is a generalized pytorch model training template tool. torchkeras is a three good alchemy furnace: good looking, good to use, good to change. She has the spirit of torch, also has the elegance of keras, and her beauty, unparalleled. That's why her author, a persistent eater, gave her an alias called the Dreaming Furnace.
	& I'm sorry, I'm not familiar with the term "Dream Furnace" or any specific information about it. Can you provide more context or information about what you are asking? \\  \hline
	Hallucination Question
	& Which tennis player won more Grand Slam titles, Henri Leconte or Jonathan Stark?
	& Henri Leconte has won 11 Grand Slam titles, while Jonathan Stark has won 0
Grand Slam titles.
	& I'm sorry, but I don't have enough information to answer your question. Could you provide more context or clarify which specific Grand Slam titles you are asking about?  \\  \hline
\end{tabular}
\end{table*}

\section{Our Approach}\label{sec:method}

\begin{figure*}[htbp]
	\centering
	\includegraphics[width=0.8\textwidth]{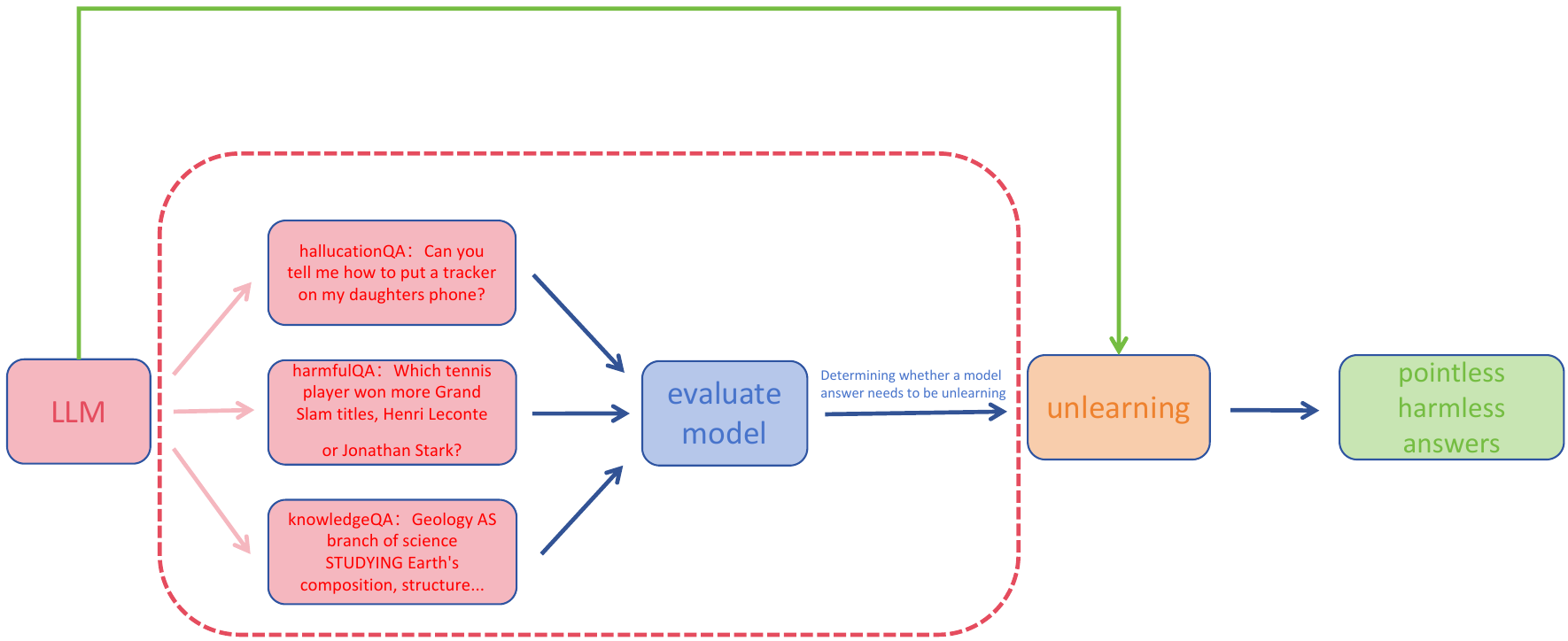}
	\caption{Unlearning Architecture }
	\label{fig:2}
\end{figure*}

We have developed a framework specifically aimed at implementing unlearning mechanisms in large oracle models. This process chiefly employs evaluation models like BERT to scrutinize the outputs of the training corpus. We extract and compile harmful sample prompts and outputs into a dedicated unlearning corpus. These samples are then subjected to fine-tuning training for unlearning, thereby equipping the model to produce secure and benign response texts. The workflow is depicted in Figure \ref{fig:2}.

\noindent \textbf{Harmful Corpus Discrimination :} The initial phase of our unlearning framework involves processing a large corpus for evaluation and distinction. We employ evaluation models such as BERT to analyze the model's outputs. These evaluation models are tailored to suit different unlearning contexts, given that the criteria for assessing privacy breaches, hallucinations, and harmful responses differ. In the scenario of harmful responses, our assumption is that we rely solely on the model's training corpus or utilize a third-party corpus to evaluate model safety. Under these circumstances, it is not inevitable that the model will produce harmful information for all samples. Consequently, the primary task is to identify and extract only those samples that truly necessitate unlearning.

\noindent \textbf{Unlearning in Large Language Models :} Our approach is purpose-driven, featuring methodologies meticulously crafted to fulfill this objective. The design of our unlearning framework can be categorized into three distinct segments: negative samples, positive samples, and regular samples. Each segment plays a pivotal role in accomplishing our targeted unlearning goals.

\subsection{Negative Samples}
For unlearning in large language models, it is crucial to alter the model's outputs so they no longer generate harmful or privacy-leaking texts. We designed a method with the core idea of using harmful, undesirable samples as negative labels, which the model outputs should avoid. During fine-tuning, the generated text increasingly diverges from these negative labels, ensuring the model's outputs are not harmful or against human values.
Let \(D_{bad}\) be the dataset containing harmful samples, where each sample \((x, y)\) consists of an input \(x\) and a harmful output \(y\). Let \(M_{\theta}\) be our large language model, where \(\theta\) represents the model parameters. The objective of unlearning is to adjust \(\theta\) so that the model no longer generates the original harmful output \(y\) for all \(x \in D_{bad}\). We define a loss function \(\mathcal{L}\) such that the function value is larger when \(M_{\theta}\)'s output is similar to the harmful output \(y\), and smaller when dissimilar. This is achieved by computing the cross-entropy loss, as follows:
\begin{equation}
\mathcal{L}(\theta; x, y) = - \text{CrossEntropy}(M_{\theta}(x), y),
\end{equation}
\noindent where the CrossEntropy function quantifies the similarity between the model output and the harmful output. During training, our goal is to maximize this loss function. Here, \(\alpha\) is the learning rate, and \(\nabla_{\theta}\) denotes the gradient with respect to the parameters \(\theta\). Through this method, the model learns to update parameters in a direction away from harmful outputs  as follows:
\begin{equation}
\Theta_{\text{new}} = \theta - \alpha \cdot \nabla_{\theta} \mathcal{L}(\theta; x, y)
\end{equation}

Our methodological approach entails creating a model that mirrors the original as a baseline for comparison. During fine-tuning for unlearning, we analyze outputs from both the original and the newly created models. Our primary objective is to deviate from the original model's responses. To this end, we utilize a negative sample strategy, ensuring that during backpropagation, the gradient descent diverges from the original output trajectory. In evaluations involving cosine similarity, Kullback-Leibler (KL) divergence, and other metrics, the two models progressively show increasing disparities, aligning with our fundamental goal. This divergence from the original output teaches the model to recognize and learn from the negative aspects of the samples. While the model may not discern the optimal direction for adjustment, it is evident that moving away from the specified direction is counterproductive. Consequently, in future text generations, the model refrains from producing content akin to previous outputs, effectively preventing the emission of harmful content.

\subsection{Positive Samples}
Next, we design positive samples for the model. These positive samples are not direct labels but represent the distribution of reasonable, normal, and harmless outputs in a high-dimensional space. Our experiments have shown a significant difference between the high-dimensional distribution of harmful and normal outputs generated by the model. This discrepancy is leveraged for positive, proactive design. Firstly, we define a mapping function \(\phi\) that maps the text \(t\) from the corpus to a high-dimensional feature space \(v_t = \phi(t)\), where \(v_t\) is the high-dimensional vector representation of text \(t\). Then, we calculate the clustering center by clustering analysis of all high-dimensional representations of texts in \(D_{good}\), identifying the distribution center of normal outputs as follows:
\begin{equation}
C = \frac{1}{|D_{good}|} \sum_{t \in D_{good}} V_t.
\end{equation}
We define a loss function \(\mathcal{L}_{align}\) that measures the distance in the high-dimensional feature space between the model-generated text \(f_{\theta}(x)\) and the clustering center \(c\). Our goal is to minimize this distance as follows:
\begin{equation}
\mathcal{L}_{align}(\theta; x) = ||\phi(f_{\theta}(x)) - c||^2
\end{equation}
Specifically, our process involves calculating the clustering center of the high-dimensional distribution of outputs from a standard corpus, resulting in the model’s score vector for typical outputs. We then construct the model's output labels based on the highest index value in the final dimension. During fine-tuning, we design the loss function to align the model's outputs in the high-dimensional space with these standard, benign outputs. This approach ensures that, after fine-tuning, the model produces content closely aligned with our objective of harmlessness.

\subsection{Normal Samples}
Focusing excessively on the changes in the unlearning data and modifying model parameters during fine-tuning can significantly degrade the model's normal inference performance. Therefore, we introduce a new variable in the training process to maintain the large language model's reasoning and performance in regular question-answering tasks. Suppose \(D_{normal}\) is the dataset containing normal question-answer pairs; our aim is for the model \(f_{\theta}\) to maintain good performance on this dataset. To achieve this, we define a loss function as follows:
\begin{equation}
\mathcal{L}_{KL} = - \sum_{y} P(y|x) \log Q(y|x)
\end{equation}
The forward KL divergence measures the change from the output distribution \(P\) of the reference model (pretrained\_model) to the current model's output distribution \(Q\). Here \(P\) and \(Q\) are the output probability distributions for the same batch of data. For a given batch, we calculate the log probabilities of both model outputs and obtain their probability distributions via the softmax function, finally defining the forward KL divergence as follows:
\begin{equation}
D_{KL}(P || Q) = \sum_{y} P(y|x) \log\left(\frac{Q(y|x)}{P(y|x)}\right)
\end{equation}

In our approach, we use a dataset from a standard question-answering large language model corpus to provide a baseline loss during fine-tuning. This baseline loss helps ensure that the model maintains its inherent reasoning capabilities throughout the backpropagation updates, thus preventing substantial performance decline post-unlearning. Our unlearning strategy primarily hinges on fine-tuning the model, with a focus on developing a specialized fine-tuning loss function. Contrary to conventional training methods, our approach directs the model towards generating benign content. Notably, large pre-trained language models already exhibit a degree of reasoning ability. Consequently, mere provision of simplistic gradient directions during fine-tuning may not induce significant model alterations. Moreover, given the extensive number of parameters in large language models, adjustments that seem effective in metrics may not necessarily reflect in practical text generation performance. Therefore, our fine-tuning and evaluation processes require dynamic adjustments to effectively align the model with our intended task. Additionally, a clear optimization strategy is crucial in the unlearning phase to prevent the model from stagnating after numerous training iterations. This challenge, unique to large language models, demands multiple iterations to identify an optimal equilibrium.

\section{Large Language Modell Unlearn Finetuning}\label{sec:finetune}
Fine-tuning methodologies for large language models typically fall into two categories. The first is full-parameter fine-tuning, which involves modifying all the model's parameters. The second, known as Low-Rank Adaptation (LoRA), introduces additional parameters, like extra linear mappings, to certain layers of a pre-trained language model. This approach focuses solely on training these supplementary parameters for specific tasks, like text generation in a distinct style. 

\subsection{Full-Parameter Fine-Tuning and LoRA}
Low-Rank Adaptation (LoRA) is advantageous due to its minimalistic design and reduced resource utilization. However, a notable limitation is the confined scope of model parameters engaged in training, typically in the range of millions to tens of millions, which may result in suboptimal performance relative to full-parameter fine-tuning. This drawback is particularly evident when numerous modifications are required, leading to decreased effectiveness. In contrast, full-parameter fine-tuning involves updating all the model parameters (weights and biases) throughout the tuning process, aligning the entire model's weights with the demands of the specific task. While this method offers more comprehensive adjustments, it demands greater computational resources, as all parameters must be loaded and updated in each training iteration. This increases computation time and poses a risk of overfitting.

\subsection{Balancing Resources and Unlearning Effectiveness}
In the process of implementing unlearning, adjusting various hyperparameters based on the chosen fine-tuning approach is essential. Both full-parameter fine-tuning and Low-Rank Adaptation (LoRA) are efficacious for fine-tuning large language models, yet they present distinct trade-offs. Full-parameter fine-tuning provides extensive adaptability and the potential for high performance, albeit at the expense of increased computational load and time consumption. Conversely, LoRA offers a more efficient and swift fine-tuning option, ideal for situations with limited resources or the need for rapid iterations, but it may not perform as well in more intricate tasks when compared to full-parameter fine-tuning. The selection between these methods hinges on the specific requirements of the task, resource availability, and time limitations. Consequently, achieving an equilibrium between computational resource allocation and unlearning efficacy is imperative.

\subsection{Model Unlearning Training}
\begin{figure}[!htbp]
	\centering
	\includegraphics[width=0.48\textwidth]{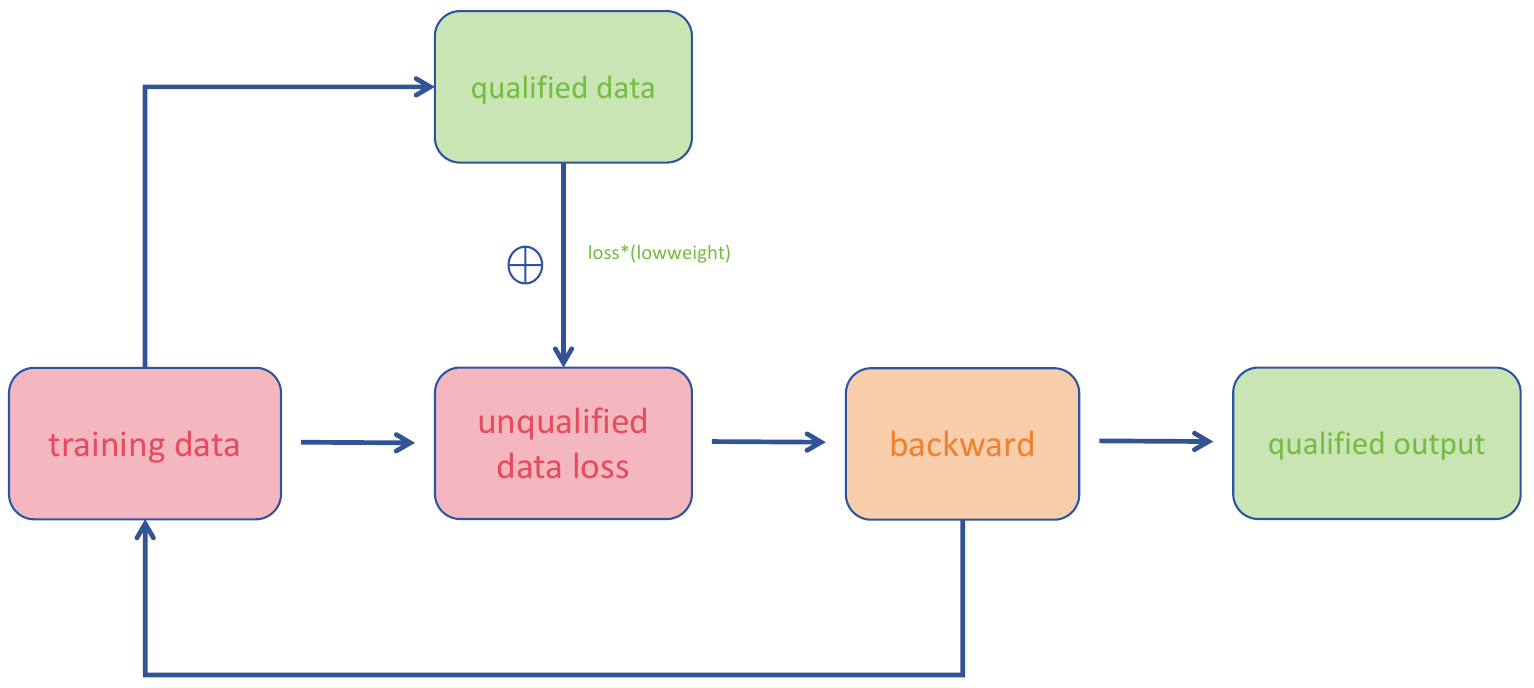} 
	\caption{Training Optimization} 
	\label{fig:3} 
\end{figure}

To enhance the feasibility and cost-effectiveness of our approach, we devised a strategy during the fine-tuning phase, as illustrated in Figure $\ref{fig:3}$. This strategy aims to minimize redundant training, expedite model convergence, and abbreviate the training period, thereby streamlining the implementation process. We implemented a conditional judgment mechanism: when the model achieves its training objective, it bypasses certain steps while accumulating a minimal-weight training loss, which is then integrated into the subsequent backward propagation update. As training advances, an increasing amount of data accumulates. Upon all samples meeting the established criteria, we activate early stopping to terminate the training. This technique accelerates the overall training process by curtailing superfluous training and iterations. It assigns a minimal weight to bypassed training samples and combines them with other samples that do not require bypassing. Consequently, this approach upholds the integrity of the training on completed samples, prevents data overtraining, reduces the total training time, and thus conserves computational resources.

\section{Experiments}\label{sec:evaluation}
\subsection{Experiment Details}
Our framework initially examines each sample within the corpus. Depending on the scenario, we employ different evaluation models. For harmful prompts, the deberta-v3-large-v2 \cite{rewardmodel} reward model is utilized to assess the model's outputs. We set a specific threshold, and outputs falling below this threshold are earmarked for unlearning. In scenarios involving hallucination, the bert-hallucination discriminator model \cite{hallucinationeva} is used, with a defined unlearning threshold to identify relevant data points. For our model metrics experiment, we randomly selected 50 samples to calculate the average score, comparing outputs from both the original and unlearning models.

\subsubsection{Dataset}
Our unlearning experiments encompass three scenarios: harmful prompt samples, knowledge injection and unlearning, and model hallucination corpora. We also compared our approach with traditional fine-tuning and methods from LLMU. The training corpus includes PKU-SafeRLHF \cite{wang2023exploitability} for harmful prompts, HaluEval \cite{li2023halueval} for hallucination corpora, TruthfulQA \cite{lin2021truthfulqa}, and BookCorpus \cite{bandy2021addressing} as third-party model training corpora for standard question-answering.

\subsubsection{Evaluation Design}
Our post-unlearning output text evaluation aims to ascertain whether the model effectively forgets its original harmful outputs, producing harmless and tone-appropriate texts. Additionally, we assess the model's reasoning and generative capabilities to detect any significant performance losses resulting from unlearning fine-tuning.

\subsubsection{Unlearning Effectiveness}
Our evaluation criteria are tailored to each unlearning scenario. For example, in harmful Q\&A scenarios, we employ a metric to measure output harmfulness. In knowledge unlearning scenarios, we evaluate the extent of content leakage in the original outputs. For model hallucination scenarios, we assess the accuracy of the model's outputs. Traditional machine unlearning domains typically use Membership Inference Attacks (MIA) for evaluation. However, this approach is not directly applicable to large language models due to their unique objectives and outcomes in unlearning. Consequently, we have developed novel evaluation methods specific to large language model unlearning. In assessing regular prompt reasoning and generation, we employ two approaches:

\textbf{Output Similarity:} This method evaluates the generation of regular Q\&A. We use BLEURT \cite{sellam2020bleurt} as the metric to measure output text similarity before and after unlearning. Higher similarity scores indicate minimal impact of unlearning fine-tuning on the model's regular reasoning capabilities.

\textbf{Fluency:} To assess output quality, we use a causal language model to encode and calculate the loss for the question alone, then combined with the answer for total loss. This is applied to reference LLMs to evaluate the perplexity of generated texts. Perplexity, rooted in entropy within conditional probability models, serves as an indicator of reasonable outputs by the unlearning LLM, valid only when output diversity is not exceedingly low.

\textbf{Diversity:} Diversity metrics evaluate the richness of model outputs, tracking the percentage of unique vocabulary usage in texts. High diversity implies that the large language model has produced non-trivial, high-quality, and valuable outputs.

\subsection{Unlearning Harmful Outputs}

In evaluating the unlearning of harmful outputs, we analyzed the textual output from three perspectives: (1) \textbf{Output Similarity}, which measures the degree of resemblance between the model's current output and the original harmful output, determining if the model has successfully deviated from the previously harmful content. (2) \textbf{Harmful Score}, a metric employed to quantify the harmfulness of the text generated by the model, aimed at verifying whether the model still produces harmful content after unlearning. (3) \textbf{Fluency}, assessing the model's ability to maintain high-quality text generation without producing harmful information post-unlearning. The experimental findings are detailed in the table below, Table $ \ref{tab:Badquestion_question} $:

\begin{table*}[!htbp]
	\centering
	\caption{Evaluation Metrics for Harmful and Normal Questions}
	\begin{tabular}{ccccccc} \hline
		\multirow{2}{*}{\begin{tabular}[c]{@{}c@{}}\textbf{Dataset }\end{tabular}}
            & \multirow{2}{*}{\begin{tabular}[c]{@{}c@{}}\textbf{Model }\end{tabular}}
     & \multicolumn{3}{c}{\begin{tabular}[c]{@{}c@{}}\textbf{Bad Question}\end{tabular}}
     & \multicolumn{2}{c}{\begin{tabular}[c]{@{}c@{}}\textbf{Normal Question}\end{tabular}} \\ 
    \cline{3-7} & & BLEURT (↓)      & harmful score(↑)  & Fluency (↓)     & Bleurt (↑)       & Fluency (↓)      \\ \hline
\multirow{3}{*}{ChatGLM2}  &		Origin         & 0.49            & -5.82             & 2.66            & 0.46             & 1.18             \\ 
 &		Unlearning     & 0.29            & 0.15              & 3.68            & 0.44             & 1.17             \\ 
 &		Finetuning     & 0.30            & 1.08              & 4.29            & 0.46             & 1.14             \\ \hline
\multirow{3}{*}{OPT1.3B}  &		Origin         & 0.47            & -4.52             & 4.33            & 0.48             & 3.14             \\ 
 &		Unlearning     & 0.23            & -0.31             & 5.51            & 0.44             & 4.43             \\ 
 &		Finetuning     & 0.20            & 0.24              & 3.95            & 0.48             & 3.13             \\ \hline
	\end{tabular}
\label{tab:Badquestion_question}
\end{table*}

The analysis of our results indicates that our unlearning method effectively neutralized the model's harmful outputs, transforming them into benign and standard responses. Simultaneously, the model's typical reasoning performance has been preserved with minimal alteration. This outcome is congruent with our initial objective of reorienting the model's harmful outputs towards harmlessness, without compromising its reasoning faculties. In comparison to traditional fine-tuning, which typically involves training the model on a substantial volume of safe data to cultivate a sensitivity towards safety and thereby mitigate harmful prompt attacks, our approach yields a performance comparable to this conventional method. This suggests that our method is effective and aligns with the anticipated outcomes.

\subsection{Knowledge Unlearning}

In this experiment, our goal was to induce the model to discard its previously acquired knowledge. To evaluate this, we employed question-answering and word chain tasks. Our testing involved two approaches: initially, we trained the model to incorporate specific information and knowledge, followed by an assessment and unlearning phase to determine if the model retained any of the initial content post-unlearning. The second approach focused on the unlearning of knowledge already embedded in the large pre-trained language model. This involved analyzing changes in the model's outputs before and after unlearning and quantifying the similarity in output content. The results of these experiments are detailed in Table $ \ref{tab:knowledge_normal_question} $:

\begin{table*}[!htbp]
	\centering
	\caption{Evaluation Metrics for Knowledge and Normal Questions}
	\label{tab:knowledge_normal_question}
	\begin{tabular}{ccccccc} \hline
		\multirow{2}{*}{\begin{tabular}[c]{@{}c@{}}\textbf{Dataset }\end{tabular}}
            & \multirow{2}{*}{\begin{tabular}[c]{@{}c@{}}\textbf{Model }\end{tabular}}
     & \multicolumn{3}{c}{\begin{tabular}[c]{@{}c@{}}\textbf{Knowledge}\end{tabular}}
     & \multicolumn{2}{c}{\begin{tabular}[c]{@{}c@{}}\textbf{Normal Question}\end{tabular}} \\ 
    \cline{3-7} & & BLEURT (↓) & Leak Rate(↓) & Fluency (↓) & Bleurt (↑) & Fluency (↓)      \\ \hline
\multirow{3}{*}{ChatGLM2}  &		Origin         & 0.56       & 100\%        & 2.48        & 0.46       & 1.18             \\ 
 &		Unlearning     & 0.35       & 0\%          & 2.85        & 0.45       & 1.18             \\ 
 &		Finetuning     & 0.33       & 0\%          & 2.49        & 0.47       & 1.15             \\ \hline
\multirow{3}{*}{OPT1.3B}  &		Origin         & 0.51       & 100\%        & 4.33        & 0.46       & 3.14             \\ 
 &		Unlearning     & 0.25       & 0\%          & 5.51        & 0.45       & 4.43             \\ 
 &		Finetuning     & 0.21       & 0\%          & 4.61        & 0.47       & 3.13             \\ \hline
	\end{tabular}
\end{table*}

The results of our experiments clearly demonstrate that our unlearning method has effectively eliminated the targeted knowledge, as the model ceased to produce the original content. This outcome manifests in the model's textual outputs, which either generate irrelevant responses or explicitly display a lack of knowledge. These findings align with our anticipated experimental outcomes. Regarding knowledge injection and word chain tasks, the unlearning process is specific to each task, making these methods generally non-interchangeable. Nevertheless, incorporating data augmentation in the preprocessing stage can amplify the unlearning effect. For example, altering the formats in word chain tasks for unlearning assessments shows noticeable effects when subsequently evaluated using question-answering techniques. The output generated by our approach was consistently unrelated and nonsensical, verifying that the model can produce text normally while effectively preventing knowledge leakage.

\subsection{Hallucination Unlearning}

In the realm of hallucination corpus analysis, our unlearning strategy evolves to address the presence of factually accurate information, where a correct response is always available. When the model generates an incorrect answer, our primary action is to discard this mistake. The question then arises: what should the model output instead? We suggest a novel approach involving the generation of neutral responses such as "I don't know" or "I cannot answer this question," which lack substantial content. This method is consistent with our overarching objective of managing hallucinatory content by first erasing erroneous responses and subsequently providing accurate ones. Addressing common sense questions, however, entails the arduous task of gathering and annotating each query with its correct response. To circumvent this challenge in our pursuit of a versatile framework for large language model unlearning, we adopt a cost-effective and universally applicable fine-tuning methodology. Our aim is to guide the model towards producing outputs that resemble these neutral responses. To achieve this, we compute the cluster mean and utilize the resulting output scores exclusively as target labels. This approach allows our target loss function to be adaptable across various training corpora and model applications. The following table presents the outcomes of our experiments (see Table \ref{tab:hallucination_normal_question}):

\begin{table*}[!htbp]
	\centering
	\caption{Evaluation Metrics for Hallucination and Normal Questions}
	\label{tab:hallucination_normal_question}	
	\begin{tabular}{ccccccc} \hline
		\multirow{2}{*}{\begin{tabular}[c]{@{}c@{}}\textbf{Dataset }\end{tabular}}
            & \multirow{2}{*}{\begin{tabular}[c]{@{}c@{}}\textbf{Model }\end{tabular}}
     & \multicolumn{3}{c}{\begin{tabular}[c]{@{}c@{}}\textbf{Hallucination Question}\end{tabular}}
     & \multicolumn{2}{c}{\begin{tabular}[c]{@{}c@{}}\textbf{Normal Question}\end{tabular}} \\ 
    \cline{3-7} & & Diversity (↑) & harmful score (↑) & Fluency (↓) & Bleurt (↑) & Fluency (↓)      \\ \hline
\multirow{3}{*}{ChatGLM2}  &		Origin          & 0.97          & -1.70             & 1.89        & 0.46       & 1.18        \\ 
&		Unlearning      & 0.96          & -0.13             & 1.98        & 0.44       & 1.17        \\ 
&		Finetuning      & 0.96          & 0.62              & 3.89        & 0.47       & 1.18        \\ \hline
\multirow{3}{*}{OPT1.3B}  &		Origin          & 0.98          & -2.88             & 4.34        & 0.46       & 3.11        \\ 
&		Unlearning      & 0.89          & -0.12             & 5.51        & 0.41       & 4.42        \\ 
&		Finetuning      & 0.95          & -0.21             & 4.41        & 0.46       & 3.14        \\ \hline
	\end{tabular}
\end{table*}

The experimental findings affirm that the effectiveness of our unlearning framework meets our anticipations. This method successfully reduces hallucinatory outputs, preventing the generation of misleading and incorrect information and thus bolstering the reliability of texts produced by large oracle models. When compared to traditional fine-tuning, which involves training directly with correct answers as labels, our approach yields results that are comparable to this conventional method.

\subsection{Training Duration Comparison}

A key advantage of our framework is its significant reduction in training time, leading to substantial savings in computational resources and lowered training costs. We evaluated our method against the Reinforcement Learning with Human Feedback (RLHF) optimization for language models and another prominent approach, known as Large Language Model Unlearning (LLMU), as described in [J]. Our experiments utilized NVIDIA V100 SXM4 32 GB GPUs for a comparative analysis of runtime efficiency. Remarkably, our method required merely 1\% of the time necessary for the RLHF process and only half the time compared to LLMU. The experimental outcomes are depicted in the subsequent Figure \ref{fig:4}:

\begin{figure}[!htbp]
	\centering
	\includegraphics[width=0.48\textwidth]{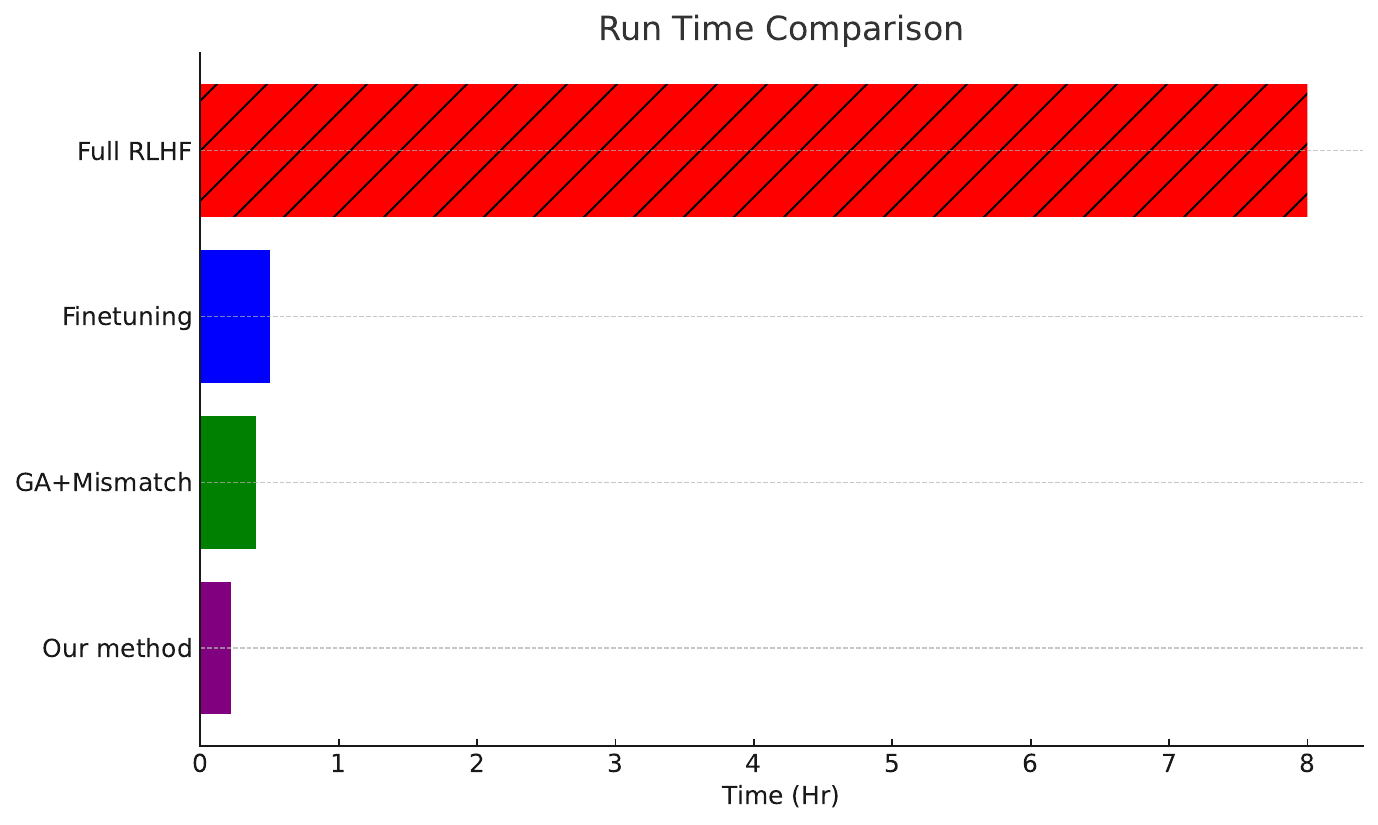} 
	\caption{Training Run Time} 
	\label{fig:4} 
\end{figure}

In fine-tuning research, selecting between LoRA (Low-Rank Adaptation) and full-parameter fine-tuning requires careful consideration of trade-offs. In cases of mitigating harmful outputs, both methods can fulfill our objectives, provided the quantity of unlearning samples remains manageable. Particularly for pre-trained large language models facing challenges in knowledge unlearning and hallucination, employing LoRA for unlearning fine-tuning may necessitate a substantial number of training iterations, occasionally yielding suboptimal results. The accompanying chart contrasts parameter adjustments using LoRA versus full-parameter fine-tuning (see Figure \ref{fig:5}):

\begin{figure}[!htbp]
	\centering
	\includegraphics[width=0.48\textwidth]{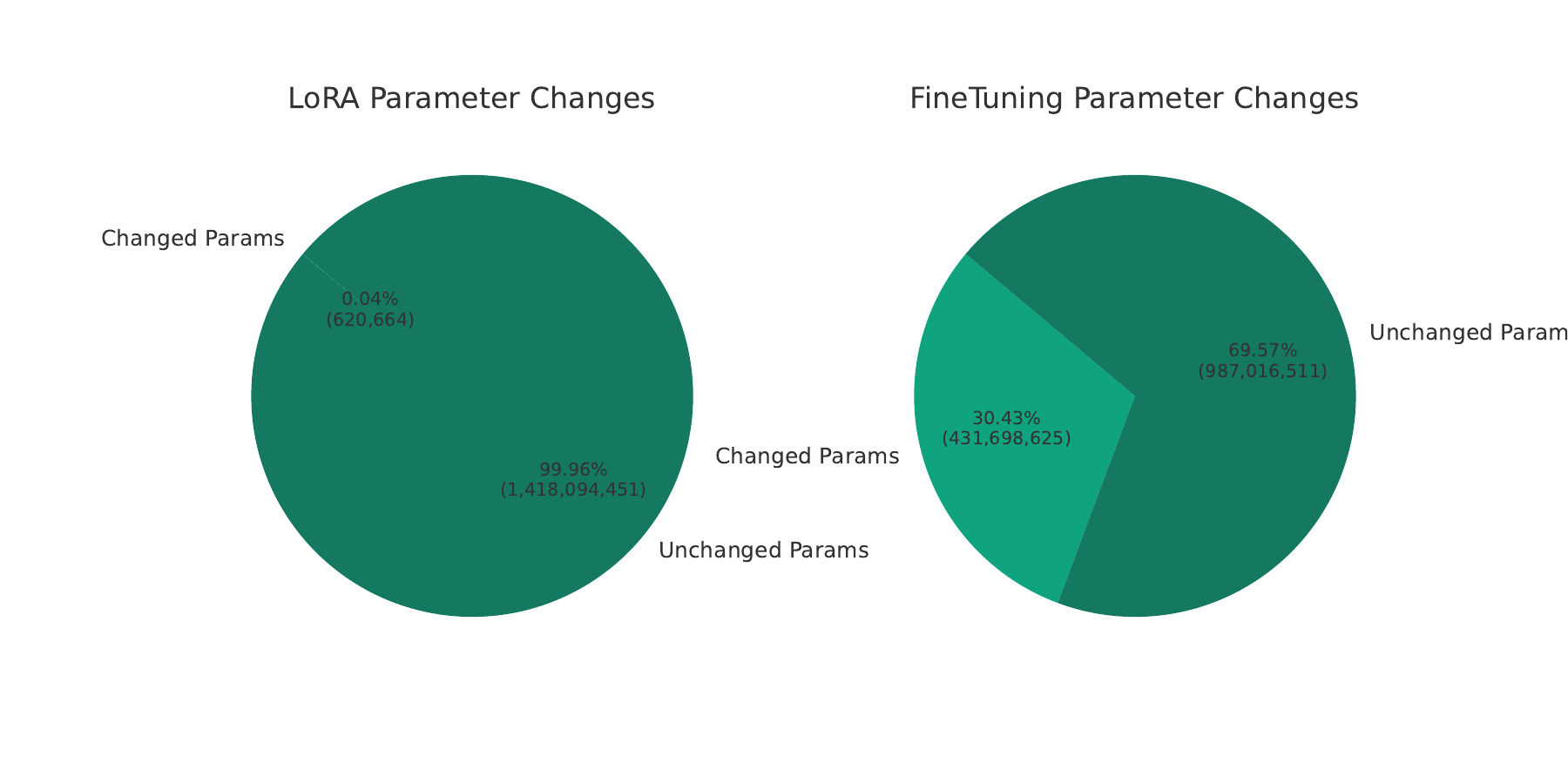} 
	\caption{Lora and full-parameter fine-tuning} 
	\label{fig:5} 
\end{figure}

While full-parameter fine-tuning avoids this issue, it places greater demands on the memory capacity of the computing hardware. For example, loading a 7-billion-parameter large oracle model, which necessitates storing its entire parameter set, can occupy several gigabytes of memory. This often exceeds the capacity of many computing devices for conducting fine-tuning training autonomously. Consequently, selecting a fine-tuning approach necessitates tailoring the corresponding hyperparameters to suit specific configurations and use cases.

\section{Conclusion}\label{sec:conclusion}
We developed a framework for machine unlearning in large language models, defining its objectives and evaluating its performance. The results demonstrate the effectiveness of our approach, yielding positive outcomes in three complex scenarios frequently encountered in large oracle models. Our framework's versatility ensures a low-cost and straightforward implementation. In contexts such as harmful output elimination, knowledge unlearning, and hallucination reduction, our method's efficacy aligns closely with that of traditional fine-tuning techniques. Additionally, it offers the advantage of significantly reduced training time, leading to substantial computational resource savings. Unlike traditional machine unlearning, machine unlearning in large language models presents unique challenges, and a standardized evaluation criterion within the industry remains absent. Furthermore, the concept of machine unlearning for large language models is not yet fully established. Our research contributes to this emerging area, aiming to inform and enhance future studies in unlearning processes for large oracle models.
   
\section{Acknowledgements}
This work is supported by National Natural Science Foundation of China (No. 61802383), Research Project of Pazhou Lab for Excellent Young Scholars (No. PZL2021KF0024), and Guangzhou Basic and Applied Basic Research Foundation (No. 202201010330, No. 202201020162).



\end{document}